\documentclass[10pt]{article}
\usepackage{cmds11}
\usepackage{natbib}
\usepackage{graphicx}
\usepackage{amsmath}
\usepackage{amssymb}
\usepackage{amsfonts}
\usepackage{bm}
\usepackage{times}
\usepackage[T1]{fontenc}

\newcommand{\sig}{\bm{\sigma}}

\title{Fast Fourier Transform computations and build-up of plastic deformation
in 2D, elastic-perfectly plastic, pixelwise disordered porous media}

\author{ {\sl F. Willot$^{1,2}$, Y.-P. Pellegrini$^1$} \\[5pt]
$^1$ Département de Physique Théorique et Appliquée, CEA, BP12, 91680
Bruyères-le-Châtel, France\\
$^2$ Laboratoire de Mécanique des Solides, \'Ecole Polytechnique, 91128 Palaiseau, France\\[5pt]
francois.willot@lms.polytechnique.fr, yves-patrick.pellegrini@cea.fr}

\begin{document}

\maketitle

\paragraph {ABSTRACT:}
Stress and strain fields in a two-dimensional pixelwise disordered system are
computed by a Fast Fourier Transform method. The system, a model for a ductile
damaged medium, consists of an elastic-perfectly matrix containing void pixels.
Its behavior is investigated under equibiaxial or shear loading. We monitor the
evolution with loading of plastically deformed zones, and we exhibit a
nucleation / growth / coalescence scenario of the latter. Identification of
plastic ``clusters'' is eased by using a discrete Green function implementing
equilibrium and continuity at the level of one pixel. Observed morphological
regimes are put into correspondence with some features of the macroscopic
stress / strain curves.


\section{INTRODUCTION}
The versatile Fast Fourier Transform (FFT) method of Moulinec, Suquet, and
Michel represented a breakthrough in the computation of the stress and strain
fields in linear or nonlinear composites. This method uses the
Lippmann-Schwinger integral equation of the strain field in a homogeneous
reference linear medium, written in the Fourier space. Nonlinearity is embedded
via a pointwise heterogeneous polarization field which depends on the
constitutive law, and which is computed in direct space. The integral equation
is solved iteratively through FFT and inverse FFT transformations for which
efficient routines are available. The ill-convergence of the basic iterative
procedure is alleviated by use of an ``augmented Lagrangian method" and by
Uzawa's algorithm, see \cite{MICH00} for details.

 \qquad With some modifications, this method
is used hereafter to investigate the build-up and incipient localization of
plastic deformation in elastic-perfectly plastic porous pixelwise disordered
systems \cite{WILL07}. Such systems consist of an on-lattice realization of a
random system where the material properties of adjacent material elements are
statistically uncorrelated from point to point, in the limit where the size of
the material element goes to zero. Other realizations of this type of disorder
(in the bond form) include random spring or resistor networks. Such systems are
particularly attractive as benchmarks for homogenization methods, since their
microcopic disorder correlation length is the smallest possible. Then, the
self-consistent linear effective medium approximation is exact to fourth order
in correlations (in the sense of diagram expansions, e.g.\ \cite{LUCK91}), as
has been shown for dielectric media or for random resistors networks (RRNs)
\cite{LUCK91}.

\section{DISCRETE GREEN FUNCTION IN FFT CALCULATIONS}
\begin{figure}
\begin{center}
      \setlength{\unitlength}{1mm}
      \begin{picture}(40,40)
        \put(0 , 0){\includegraphics[width=4cm]{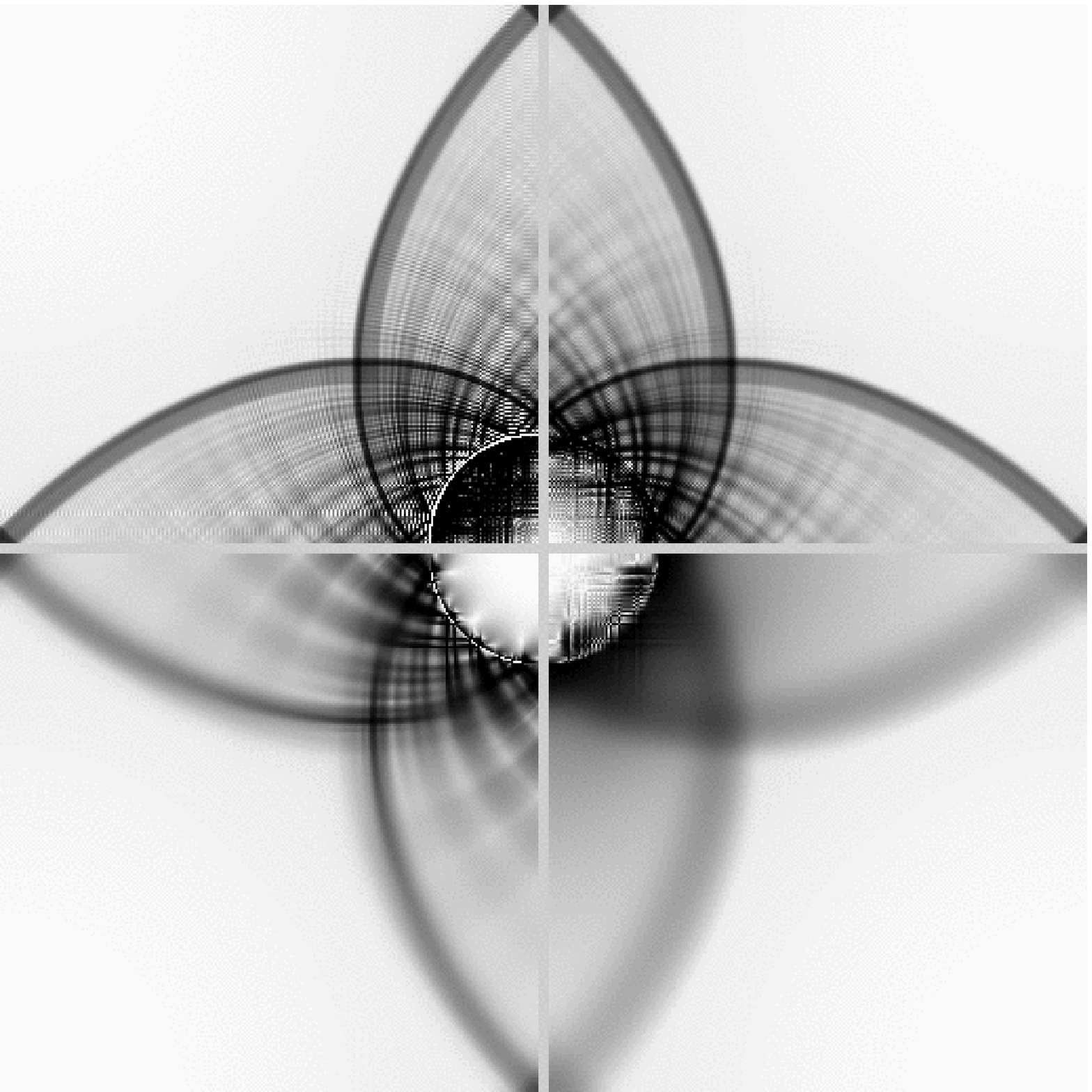}} 
        \put(2 , 2){\textbf{\textit{\large{DG2}}}}
        \put(30, 2){\textbf{\textit{\large{DG1}}}}
        \put( 2,35){\textbf{\textit{\large{CG1}}}}
        \put(30,35){\textbf{\textit{\large{CG2}}}}
      \end{picture}
      \hspace{1.5cm}
      \begin{picture}(40,40)
        \put(0 , 0){\includegraphics[width=4cm]{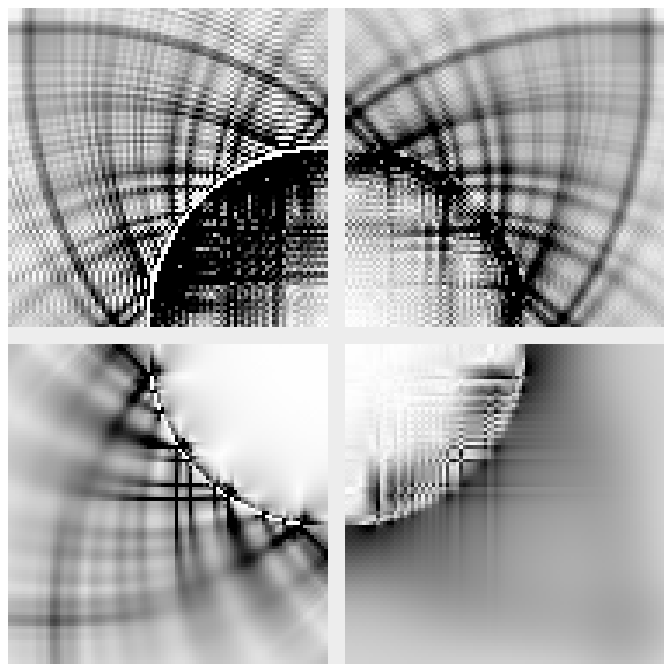}}
        \put(2 , 2){\textbf{\textit{\large{DG2}}}}
        \put(30, 2){\textbf{\textit{\large{DG1}}}}
        \put( 2,35){\textbf{\textit{\large{CG1}}}}
        \put(30,35){\textbf{\textit{\large{CG2}}}}
      \end{picture}
\caption{\label{graphEpsEq12} Differences between the methods. Periodic case
with one void under equibiaxial loading. The gray field shows the equivalent
strain field norm, $\varepsilon_{\rm
eq}=[(2/3)\varepsilon_{ij}'\varepsilon_{ij}']^{1/2}$, where
$\varepsilon_{ij}'=\varepsilon_{ij}-\delta_{ij}\varepsilon_{kk}/3$. Pixels near
the void surface with the highest field values have been thresholded out ($\sim
1.5 \%$ of the total number). The right image is an enlargement.}
\end{center}
\end{figure}

A two-dimensional (2D) square lattice of pixels $(i,j)$ of size $L^2$ is
considered. Pixels are either randomly chosen as voids, in concentration $f$,
or as elastic-perfectly plastic matter elements with flow stress $Y$.
Deformation theory \cite{LUBL90} is used. Using the continuum Green tensor of
the strain (GT) as in \cite{MICH00} (we call this approach ``CG1''), we observe
that : (i) convergence of the FFT method is slow for infinite contrast; (ii)
the fields, and the plastically deformed zone where the Mises norm of the
stress $\sigma_{\rm eq}$ is locally equal to the flow stress $Y$ exhibit a
spurious ''checkerboard" pattern (see Fig.\ \ref{graphEpsEq12}) which
complicates a subsequent identification of
plastic clusters (cf.\ Sec. 3).\\

\emph{Discrete Green functions ---}  In the alternative solution considered
here, introducing the unit vectors along the axes $\mathbf{\hat e}^i$ such that
$e^i_j=\delta_{ij}$ (the Kronecker symbol), we follow the RRN scheme of Ref.\
\cite{LUCK91} and use forward and backward finite-difference schemes for the
compatibility and equilibrium equations (discrete Fourier Transforms in
elasticity are also used in Ref.\ \cite{DREY99}):
\begin{equation}
\label{eq:epsij}
\varepsilon_{ij}(\mathbf{x})=\frac{1}{2}\left[u_j(\mathbf{x}+\mathbf{\hat
e}^i)-u_j(\mathbf{x})+(i\leftrightarrow
j)\right],\quad\sum_j\left[\sigma_{ij}(\mathbf{x})-\sigma_{ij}(\mathbf{x}-\mathbf{\hat
e}^j)\right]=0.
\end{equation}
Contrary to a centered-difference scheme, equilibrium is enforced here on any
finite connex subset of pixels. In the discrete Fourier representation with
discrete Fourier momenta $q_i=2\pi m_i/L$, $m_i=0,1,\ldots L-1$ (the
conventions of Ref.\ \cite{LUCK91} are used), these equations read [here
$i=(-1)^{1/2}$]:
\begin{equation}
\label{eq:epssigk} \varepsilon_{kl}=(i/2)\left(k_k u_l+k_l u_k\right),\quad i
k_l^* \sigma_{kl}=0,\qquad\mathrm{with}\quad k_j\equiv 2 \sin(q_j/2) e^{i
q_j/2},
\end{equation}
where $*$ denotes the complex conjugate. The associated GT for the strain is:
\begin{equation}
\label{eq:gd} G_{ijkl}(\mathbf{q})=-\left(
N^{-1}_{jl}k_ik^*_k+N^{-1}_{il}k_jk^*_k
+N^{-1}_{jk}k_ik^*_l+N^{-1}_{ik}k_jk^*_l\right)/4
\end{equation}
where the acoustic tensor is $N_{ij}\equiv k^*_k\,C_{iklj}k_l$ with $C_{ijkl}$
the elastic tensor of the isotropic reference medium of Lamé moduli $\mu$ and
$\lambda$. The restriction to generalized plane strain loading (fields
independent of $z$) is obtained by setting $q_3=0$ and retaining only the
indices $i,j,k,l=1,2$. Though an analytic expression of $G$ is available
\cite{WILL07}, the inversion of $\mathsf{N}$ is most easily carried out
numerically. The continuum GT (e.g.\ \cite{MICH00}) is retrieved in the long
wavelength limit $q_1$, $q_2\ll 1$. A centered-difference scheme would instead
lead to a real GT, simply obtainable from the continuum GT by replacing its
Fourier momenta by $k_j\equiv\sin(q_j)$.

\qquad The GT (\ref{eq:gd}) does not comply with the square symmetry of the
grid, and lacks major symmetry, due to its nonzero imaginary part:
$G_{ijkl}=G^*_{klij}$. The inconsistency comes from the fact that the
equilibrium equation in (\ref{eq:epsij}) is a balance condition for forces
transmitted by \emph{bonds} linking nearest-neighbor pixels, whereas the
constitutive law used is appropriate for material points only, i.e.\ the
pixels. As a result, the forward and backward directions on a each cartesian
axis are not equivalent, which leads to asymmetric field patterns. Two easy
work-arounds are considered. The first one, called \emph{discrete Green 1}
(DG1), consists in replacing (\ref{eq:gd}) by its symmetrized version
\begin{eqnarray}
G^{(1)}_{ijkl}&\equiv&\left\{G_{ijkl}(q_1,q_2)+G_{ijkl}(-q_1,-q_2)\right.\nonumber\\
&&\left.{}+(-1)^{i+j+k+l}\left[G_{ijkl}(-q_1,q_2)+G_{ijkl}(q_1,-q_2)\right]\right\}/4.
\end{eqnarray}
This GT corresponds to no discretization scheme in direct space, but should be
interpreted as the GT of some non-local medium in which the stiffness tensor
$C_{ijkl}$ is replaced by a non-local convolution kernel with finite range of
order 1. The second way, called DG2, consists in carrying out four different
calculations on the same system up to final convergence, employing each one of
the four $G_{ijkl}(q_1,q_2)$, $G_{ijkl}(-q_1,-q_2)$, $G_{ijkl}(-q_1,q_2)$,
$G_{ijkl}(q_1,-q_2)$ in turn, an in taking the average of the four converged
strain fields as the final result. We also consider, for comparison purposes,
field pictures obtained from CG1 by means of a local average post-processing
consisting in taking 5-point averages on the current pixel and its four nearest
neighbors. We call this the CG2 method.

\qquad Setting $k^2=k_1^2+k_2^2$, the 2D displacement field is obtained from
$\varepsilon_{ij}(\mathbf{q})$ using (the mode $q=0$ is irrelevant):
$$
u_1(\mathbf{q})=-\frac{i}{k^2}\left\{k_1\left[\varepsilon_{11}-\varepsilon_{22}\right]+2k_2\,\varepsilon_{12}\right\},\hspace{1ex}
u_2(\mathbf{q})=\frac{i}{k^2}\left\{k_2\left[\varepsilon_{11}-\varepsilon_{22}\right]-2k_1\,\varepsilon_{12}\right\}.
$$
In the computations hereafter, the bulk and shear moduli are $K=1$ and
$\mu=0.4$ and $Y=0.5$. The loading is prescribed by imposing an overall strain,
which is increased until the stress reaches its flow value \cite{MICH00}.

\qquad Differences between the methods are illustrated in Fig.\
\ref{graphEpsEq12}, where the equivalent shear strain in a periodic medium
under equi-biaxial loading is displayed, in quadrants of a unit cell with one
circular void (volume fraction $f=0.1$).\footnote{The strain localization
motif, of a surprisingly rich sub-structure -- remark the bands of finite width
which bound the localization zone -- is approximately made of logarithmic
spirals leaving the void surface at angle 45${}^o$ \cite{LUBL90} as predicted
by slip-line theory (e.g., \cite{LUBL90}), though the lines are somehow
distorted and blurred by lattice effects.} Method CG2 does not suppress the
``checkerboard'' pattern of method CG1. Method DG1 blurs excessively the strain
field. The best result without ``checkerboard'' effect is obtained with DG2.
The strain field inside the void depends on the method, but
is physically irrelevant.\\

\emph{Convergence issues ---} Two convergence indicators are used. First, we
require the stress divergence (computed in Fourier representation) to be such
that $\left\langle||\textnormal{div}\sig||^2\right\rangle$
 $<$ $\eta_1^2$ $\langle\sig\rangle:\langle\sig\rangle$. Additional steps of
the iterative algorithm are then carried out until $
\left\langle\sig^{n+1}-\sig^n\right\rangle:\left\langle\sig^{n+1}-\sig^n\right\rangle$
$<$ $\eta_2^2$ $\left\langle\sig^{n+1}\right\rangle
:\left\langle\sig^{n+1}\right\rangle$ between steps $n$ and $n+1$. The overall
prescribed tolerance is specified by the pair $(\eta_1,\eta_2)$. For
benchmarking, typical values of $\eta_{1,2}$ of order $10^{-5}$, and system
sizes $L=512$, $1024$, $2048$ were considered. For both the periodic void
lattice and the pixelwise disordered medium on which we focus hereafter, method
DG2 converges faster (i.e., in fewer iterations) than CG1. The absolute
precision on the average stress, extrapolated to infinite system sizes by an
inverse power law fit of the size dependence, is then typically $10^{-3}$. For
more precise calculations, values of $\eta_{1,2}$ of order $10^{-8}$ and system
size $L=4096$ have been used (see \cite{WILL07} for details).

\section{BUILD-UP OF PLASTIC DEFORMATION IN THE POROUS MEDIUM}
Using method DG2, we investigate the link between the overall stress-strain
curve and the development of the plastic zones where $\sigma_{\rm eq}=Y$ in a
pixelwise disordered porous medium. The medium can be considered as made of
three phases: elastic (where $\sigma_{\rm eq}<Y$), porous (where $\sigma=0$),
and plastic ($\sigma_{\rm eq}=Y$). Under increasing loading, plastic zones
develop starting from the voids (around which the shear stress is larger),
grow, coalesce and eventually ``percolate'' in the system. Void or plastic
clusters, are identified (with the Hoshen-Kopelman algorithm \cite{STAU85}) as
pixel sets of same phase connected by the nearest-neighbour criterion. The
``void-plastic'' phase comprises the voids and the plastic zones. In this phase
the local tangent shear modulus is zero. Due to local unloading effects
(possibly an artefact of the -- reversible -- deformation theory), purely
plastic clusters in the vicinity of the pores can temporarily disconnect during
the first loading stages, especially under shear loading. This has bearing on
cluster counting. This was corrected by modifying the first-neighbor
connectivity rule: we furthermore connect to the nearest void, each plastic
cluster not already connected to a void. Each cluster then contains one void at
least.
\begin{figure}[h!]
\rotatebox{-90}{\includegraphics[width=4.3cm,height=7cm]{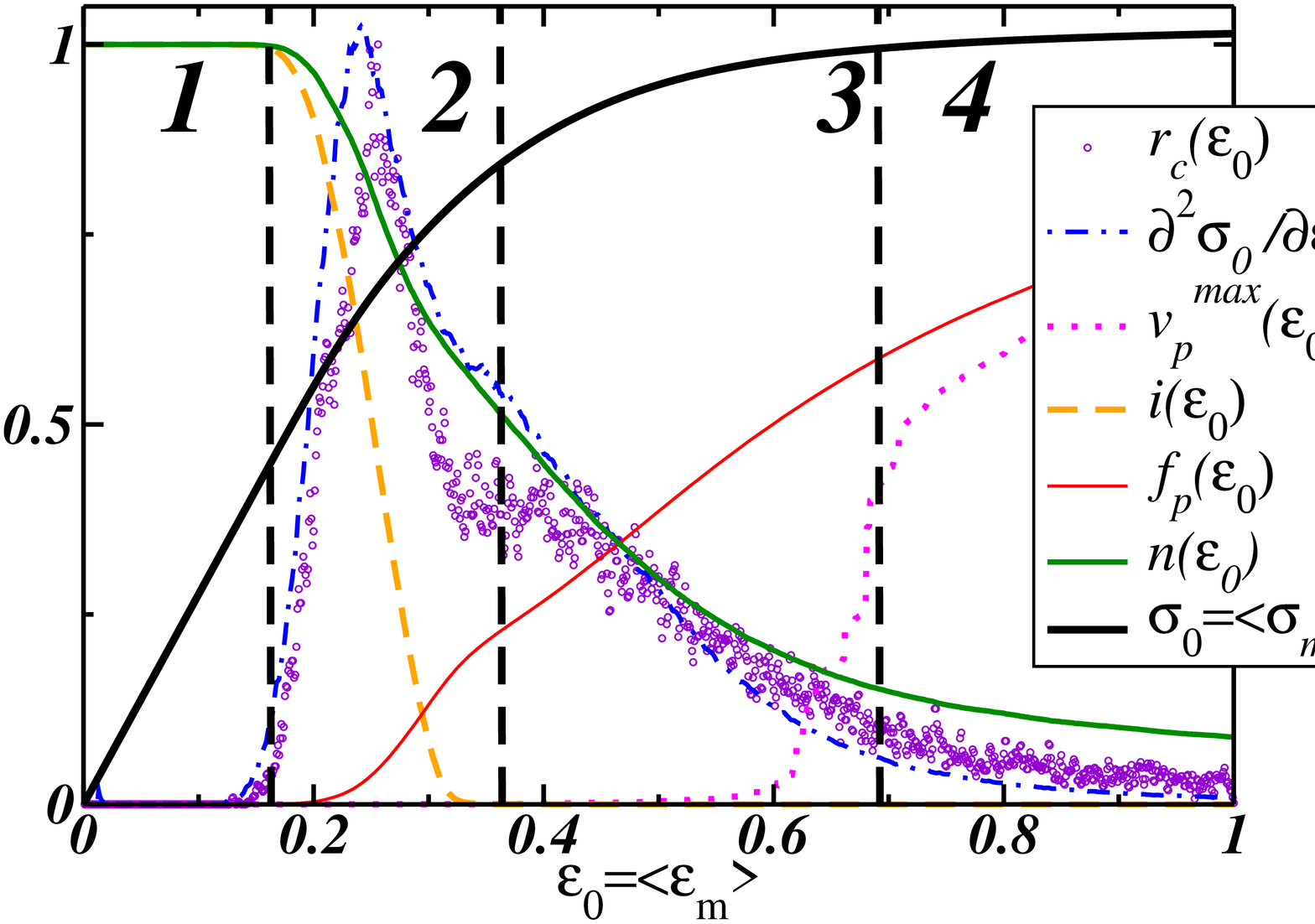}}
\rotatebox{-90}{\includegraphics[width=4.3cm,height=4.8cm]{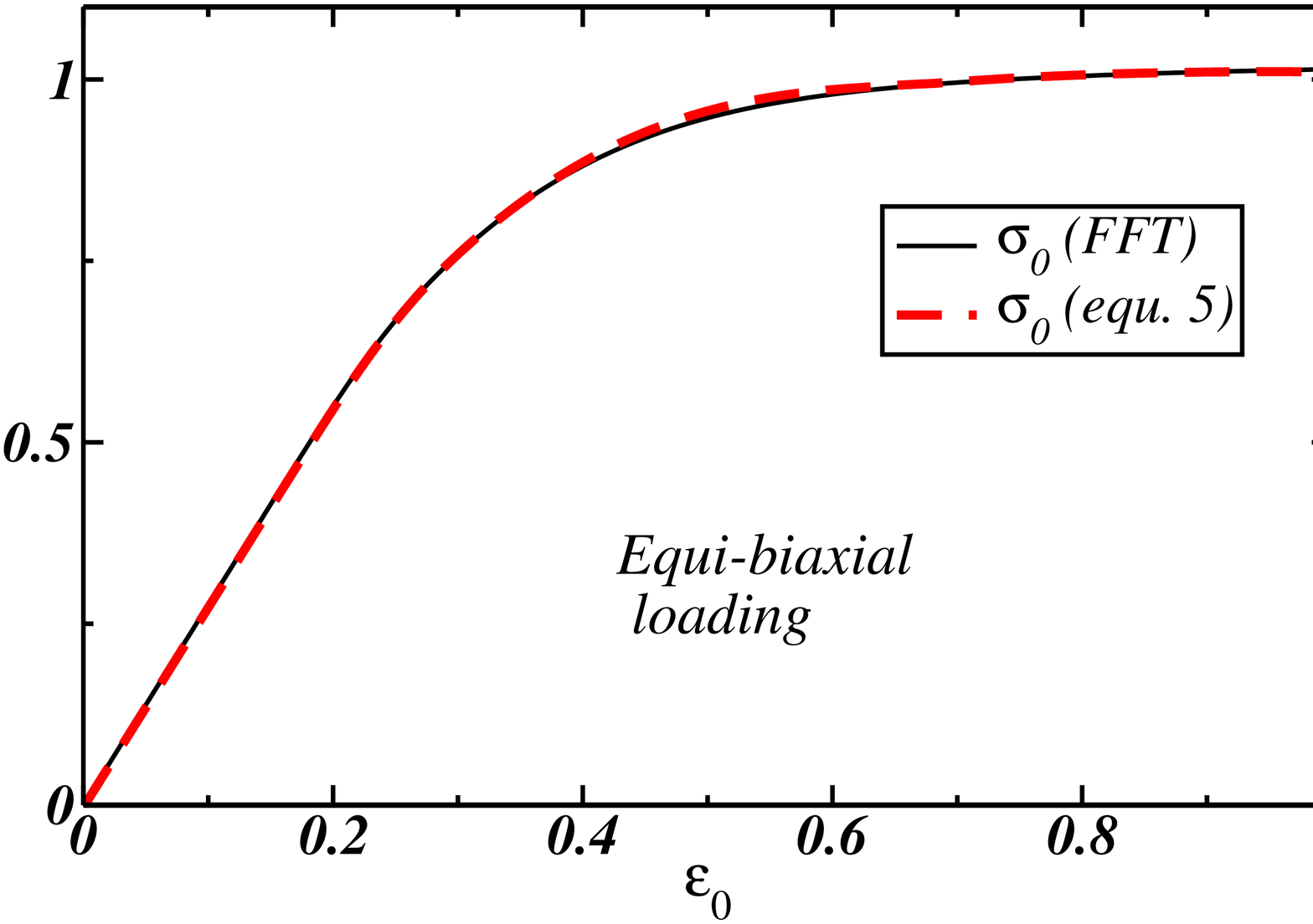}}
\caption{\label{fig:pressload} Equi-biaxial loading. Left: stress-strain curve
and geometric indicators as a function of the applied overall strain. Right:
empirical stress-loading formula compared to FFT stress/strain curve.}
\end{figure}

\begin{figure}[ht]
\begin{center}
\includegraphics[width=4.cm]{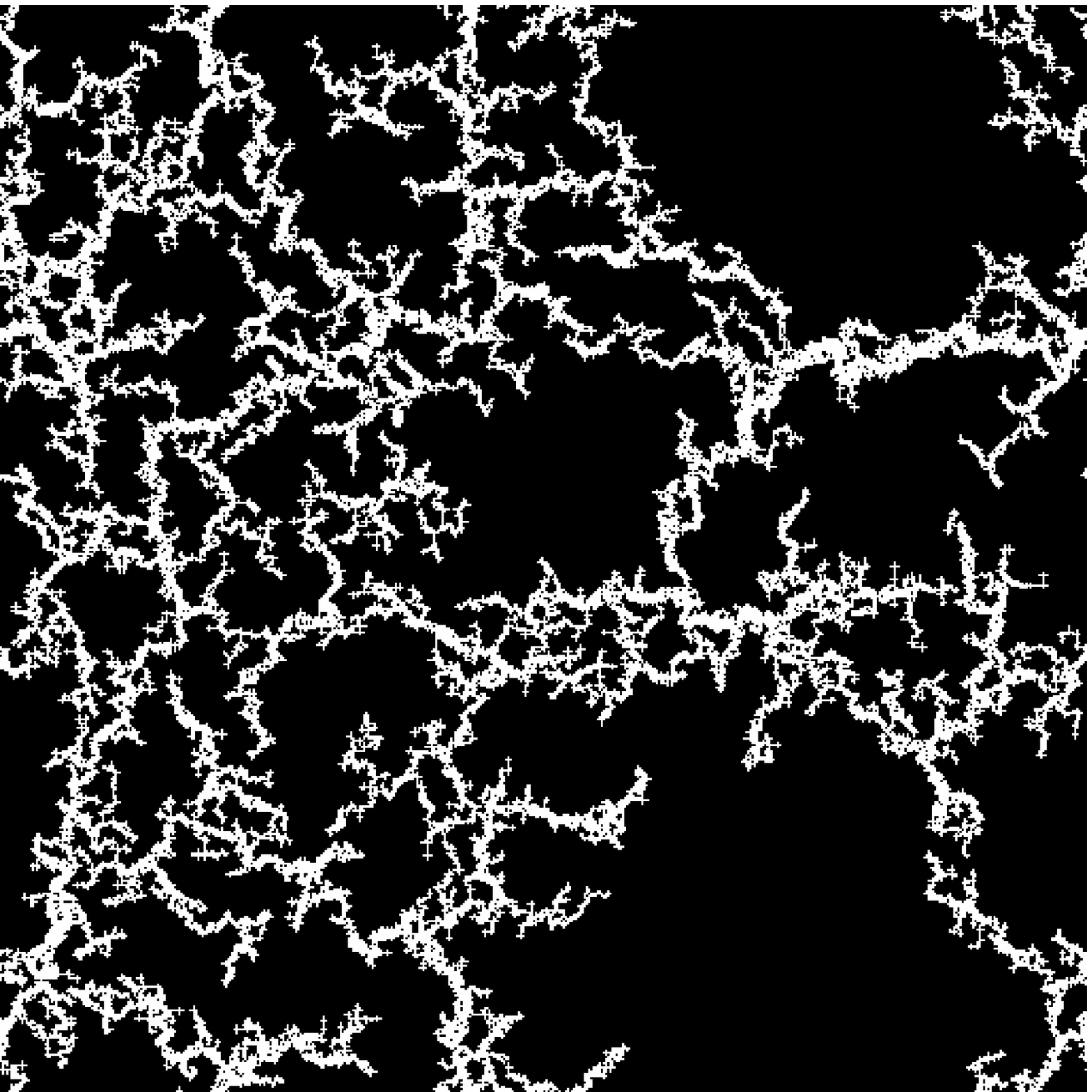}\hspace{0.5cm}
\includegraphics[width=4.cm]{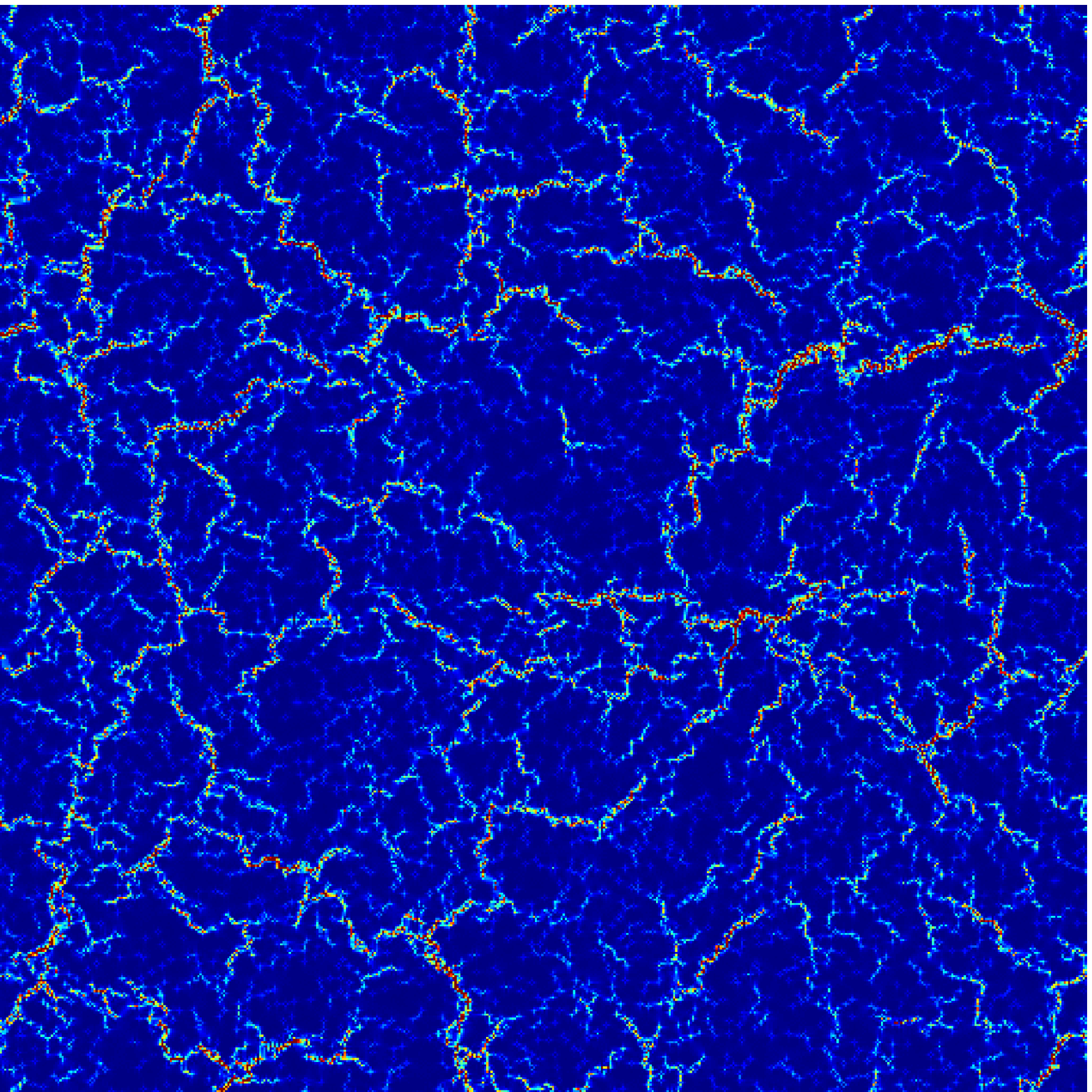}
\end{center}
\caption{\label{fig:amaspressure} Equi-biaxial loading for porosity $f=0.1$.
Left: largest void-plastic cluster (white) at percolation ($\varepsilon_0\simeq
0.9$). Right: equivalent norm of the shear strain field at a larger overall
strain ($\varepsilon_0=1.1$).}
\end{figure}

\emph{Equi-biaxial loading ---} Fig.\ 2 (left) displays, for a porosity
$f=0.01$, various geometrical indicators, as a function of the applied
equi-biaxial overall strain $\varepsilon_0=\langle \varepsilon_{\rm m}\rangle$,
along with the macroscopic equibiaxial stress/strain curve
$\sigma_0=\langle\sigma_{\rm m}\rangle(\varepsilon_0)$ (the brackets denote a
spatial average), and the opposite of its second derivative $-\partial^2
\sigma_0/\partial \varepsilon_0^2$: the normalized volume fraction of
void-plastic zone $f_{\rm p}=v_{\rm p}/(1-f)$ where $v_{\rm p}$ is the volume
fraction of void-plastic zone; the volume fraction $v_{\rm max}$ of the largest
void-plastic cluster; the proportion of isolated voids clusters
$i(\varepsilon_0)$ (i.e.\ voids not connected to a plastic zone); the number
$n(\varepsilon_0)$ of void-plastic clusters (normalized to the number of voids
at $\varepsilon_0=0$); and the coalescence rate $r(\varepsilon_0)=\partial
n(\varepsilon_0)/\partial \varepsilon_0$, multiplied by a magnifying factor so
as to make it conspicuous on the curve. Four characteristic regimes are
isolated, marked on the figure: (1) elastic loading regime: the number of
isolated voids remains constant; (2) growth of plastic zones around the voids:
the number of isolated voids diminishes, $f_{\rm p}$ grows approximately
quadratically, the coalescence rate develops a huge peak (coalescence between
plastic clusters originating from neighboring voids). The regime ends up as
$i(\varepsilon_0)=0$, all voids having developed a plastic zone; (3) regime of
stabilized coalescence: $f_{\rm p}$ increases linearly, the coalescence rate
somewhat stabilizes, then decreases more slowly as a largest void-plastic
cluster emerges; (4) stress saturation regime: it begins at the percolation of
the void-plastic zone where the largest plastic cluster grows fast, and where
$f_{\rm p}$ increases slower than linearly with the strain. From these
observations, we see that the second derivative of the stress-strain curve is
strongly correlated to the coalescence rate, and that the macroscopic flow
stress attains its final order of magnitude at percolation of the plastic zone
through the medium. Typical plastic clusters and shear strain field are
displayed in Fig.\ \ref{fig:amaspressure}. Their fractal character will be
discussed elsewhere. The shear strain is strongly localized (but not uniform)
within the plastic zones.
\begin{figure}[h!]
\rotatebox{-90}{\includegraphics[width=4.3cm,height=7cm]{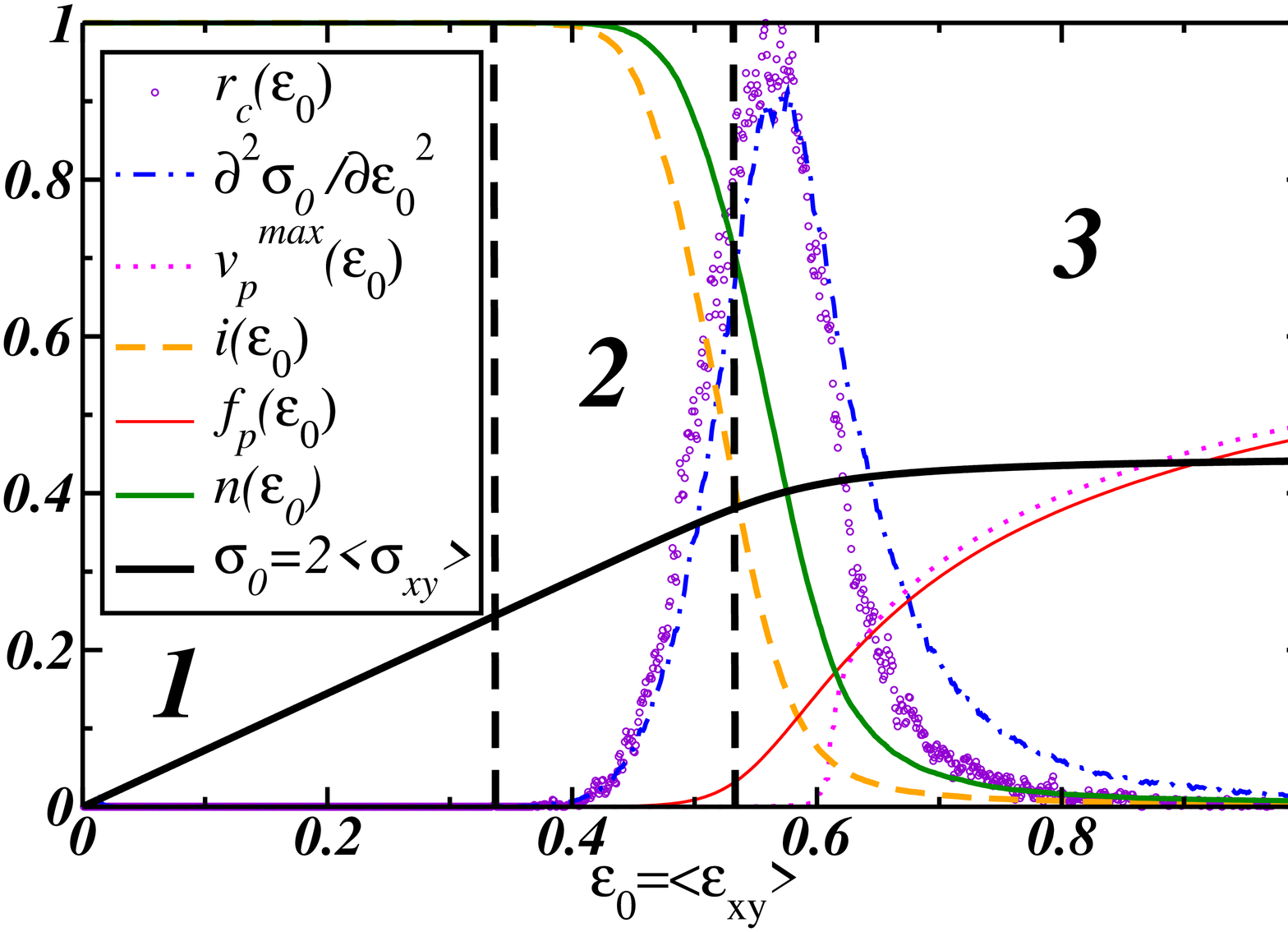}}
\rotatebox{-90}{\includegraphics[width=4.3cm,height=4.8cm]{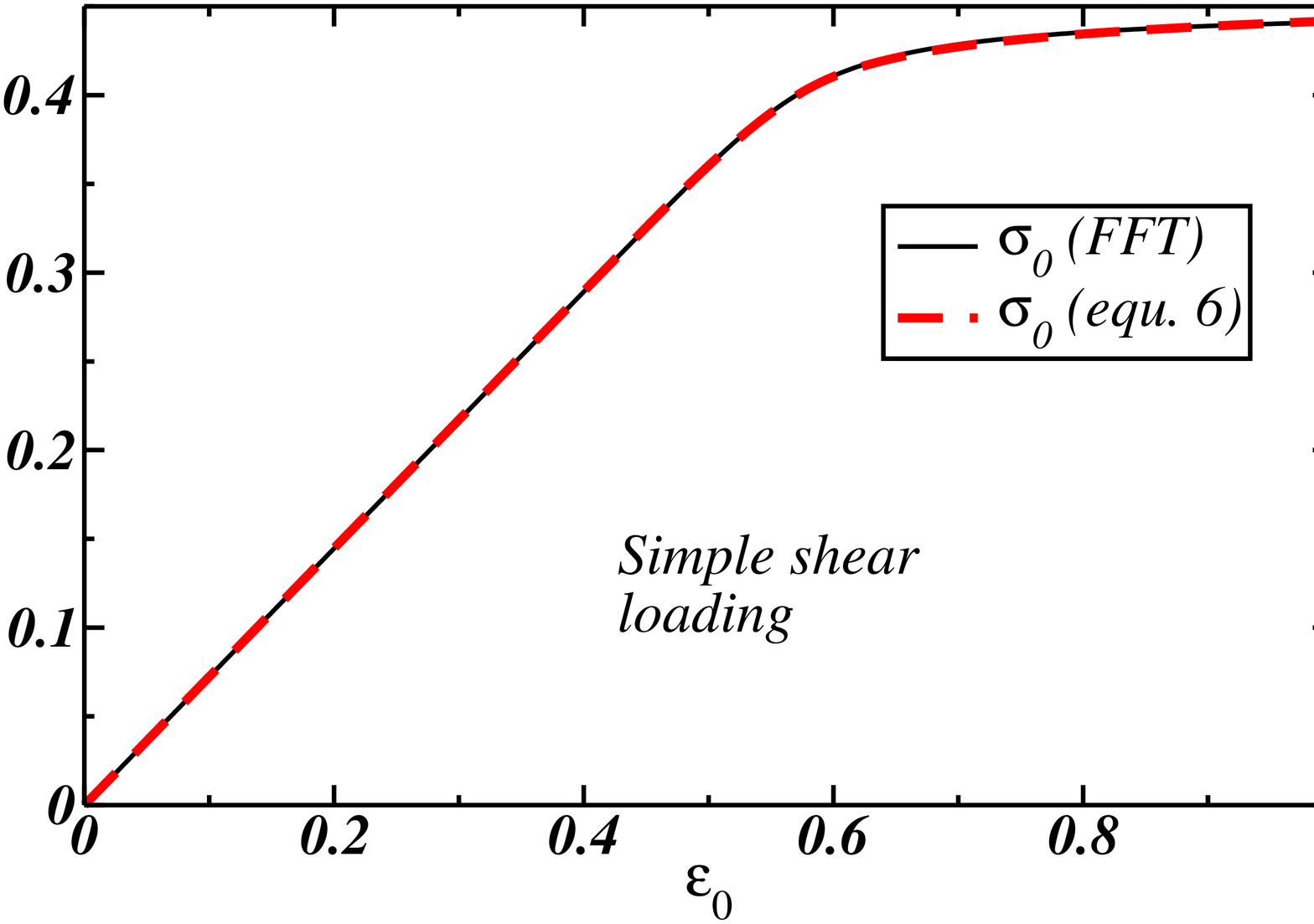}}
\caption{\label{fig:shearload} Simple shear loading. Same legend as fig. 2.}
\end{figure}
\\

\begin{figure}[ht]
\begin{center}
\includegraphics[width=3.75cm]{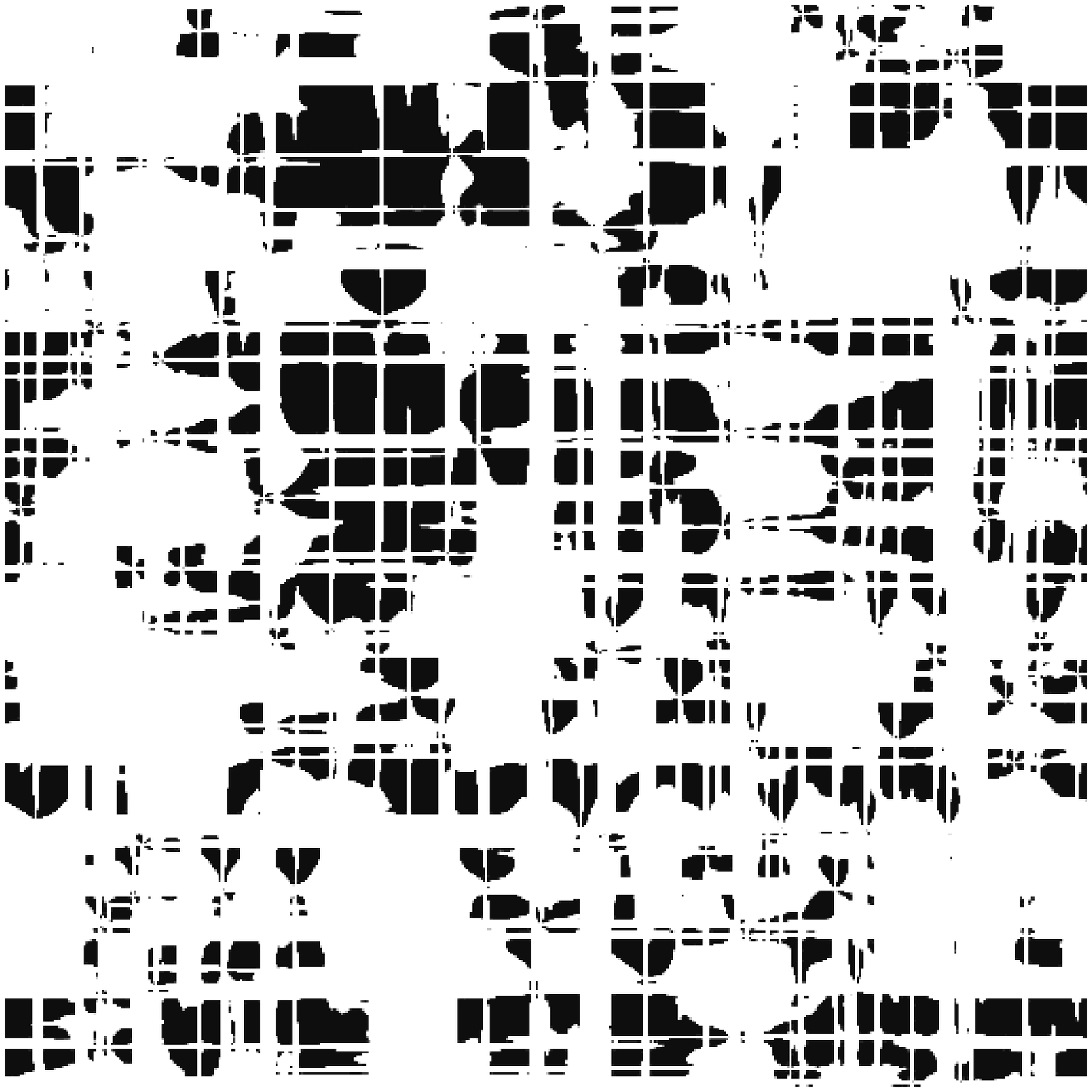}\hspace{0.5cm}
\includegraphics[width=3.75cm]{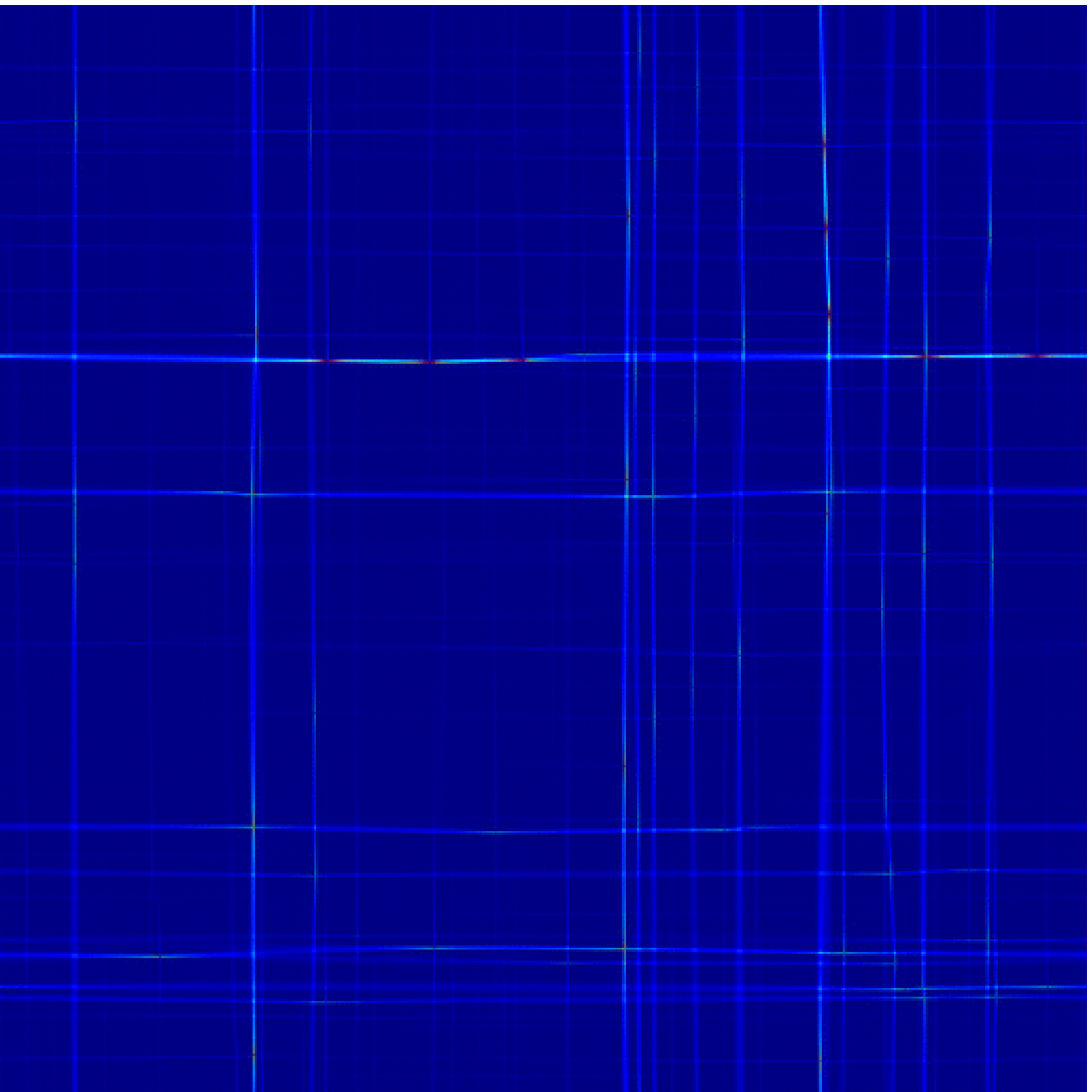}
\end{center}
\caption{\label{fig:SSnlinrandom} Simple shear loading at $f=10^{-4}$. System
size $L=1024$ at applied strain $\varepsilon_0=2$. Left: plastic (white) and
elastic (black) zones. Right: equivalent norm of the shear strain field.}
\end{figure}

\emph{Loading in simple shear ---} Fig.\ 4 (left) displays the same quantities
as Fig.\ 2, for the same medium, in simple shear loading where
$\varepsilon_0=\langle\varepsilon_{xy}\rangle$ and
$\sigma_0=\langle\sigma_{xy}\rangle$. The main difference is the disappearance
of the stabilized coalescence regime. Fig.\ \ref{fig:SSnlinrandom} displays a
typical instance of the fields for a weak porosity, with straight shear bands.
\\

\emph{Empirical formula for the stress-strain curve ---} Amusingly, one can
reproduce the stress-strain curves using the above geometric indicators and the
porosity-dependent effective compressibility and shear elastic moduli
$\widetilde{K}$ and $\widetilde{\mu}$. We indeed arrived at the empirical
formulae (the effective elastic moduli are computed on the simulated system):
\begin{equation}
\label{eq:press}\sigma_0\simeq \widetilde{K}(f)\int_0^{\varepsilon_0} {\rm
d}\varepsilon_0\,\left\{\frac{n(\varepsilon_0)-n(\infty)}{1-n(\infty)}+\alpha
\left[v_{\rm p}(\varepsilon_0)-v_{\rm p}^{\rm
max}(\varepsilon_0)\right]\right\},
\end{equation}
\begin{equation}
\label{eq:shear} \sigma_0\simeq\alpha\,f_{\rm
p}+2\widetilde{\mu}(f)\int_0^{\varepsilon_0} {\rm
d}\varepsilon_0\,\frac{n(\varepsilon_0)-n(\infty)}{1-n(\infty)},
\end{equation}
where (\ref{eq:press}) and (\ref{eq:shear}) apply to pressure and shear
loadings respectively, and where $\alpha$ is a fitting number different in each
case. In both expressions, the term containing cluster numbers reproduces the
zone of maximal curvature of the stress-strain curve (regime of maximal
coalescence rate), whereas the plastic volume fraction enters the description
of the saturation regime. These formulae are compared to the FFT stress /
strain curves in Figs.\ 2 and 3 (right). Though the present ``guesswork" should
not be taken too seriously (in particular, the dependence of $\alpha$ with
respect to $f$ has not been studied, and the plastic volume fraction does not
enter both formulas in the same manner), such an approach might nonetheless
ultimately contribute to enrich the effective-medium approach to porous media.
Indeed taking correlations into account so as to incorporate microstructural
information in the effective-medium framework is a notoriously difficult
problem. The possibility of using global indicators of geometric nature such as
the fraction of plastic zone, or cluster numbers, for which phenomenological
evolution models could be proposed might then constitute a useful alternative.
\medskip

Acknowledgments: Y.-P.P.\ thanks Pierre Suquet for drawing his attention to Ref.\ \cite{DREY99}.\\

\bibliographystyle{unsrt}

\begin{thebibliography}{99}
\bibitem{MICH00}
J.-C. Michel, H.~Moulinec and P.~Suquet.
\newblock A computational method based on augmented Lagrangians and Fast Fourier
Transforms for composites with high contrast.
\newblock {\em Comput. Model. Eng. Sci.}, 1:79--88, 2000, and references
therein.
\bibitem{WILL07}
F.~Willot.
\newblock{\em Contribution à l'étude théorique de la localisation plastique dans les poreux}.
PhD.\ thesis, \'Ecole Polytechnique, 2007 (in French).\\
\newblock http://www.imprimerie.polytechnique.fr/Theses/Files/Willot.pdf.
\bibitem{LUCK91}
J.-M.~Luck.
\newblock Conductivity of random networks: an investigation of the accuracy of
the effective-medium approximation.
\newblock{Phys.\ Rev.\ B}, 43:3933--3944, 1991.
\bibitem{LUBL90}
J.~Lubliner.
\newblock{\em Plasticity theory}.
\newblock Macmillan, New York, 1990.
\bibitem{DREY99}
W.~Dreyer, W.H.~Müller, J.~Olchewski.
\newblock An approximate analytical 2D-solution for the stresses and strains in
eigenstrained cubic materials.
\newblock{Acta Mech.}, 136: 171--192, 1999.
\bibitem{STAU85}
D.~Stauffer, A.~Aharony.
\newblock{\em An introduction to percolation theory}.
\newblock Taylor and Francis, London, 1985.
\end{thebibliography}

\end{document}